\documentclass{ws-ijmpe}
\usepackage[super,compress]{cite}
\usepackage[breaklinks]{hyperref}
\begin{document}

\markboth{Bastian B. Brandt}{The electromagnetic form factor of the pion:
Results from the lattice}

\catchline{}{}{}{}{}

\newcommand{\refc}[1]{(\ref{#1})}
\renewcommand{\vec}[1]{\mathbf{#1}}

\newcommand{\sysfullcontrol}{$\checkmark$}
\newcommand{\syshalfcontrol}{$\Box\:$}
\newcommand{\sysnocontrol}{$\times\:$}
\newcommand{\rpi}{\langle r_\pi^2 \rangle}
\newcommand{\lc}[1]{\bar{\ell}_{#1}}

\title{The electromagnetic form factor of the pion: \\ Results from the lattice}

\author{Bastian B. Brandt}

\address{Institut f\"ur theoretische Physik \\
Universit\"at Regensburg, D-93040 \\
bastian.brandt@physik.uni-regensburg.de}

\maketitle

\begin{abstract}
This review contains an overview over recent results for the electromagnetic
iso-vector form factor of the pion obtained in lattice QCD with dynamical
fermions. Particular attention is given to the extrapolation to the physical
point and an easy assessment of the control over the main systematic effects by
imposing quality criteria and an associated sign code, similar to the ones used
by the FLAG working group. Also included is a brief discussion of recent
developments and future challenges concerning the accurate extraction of the
form factor in the lattice framework.
\end{abstract}


\ccode{PACS numbers: 12.38.Gc, 13.40.Gp, 14.40.Be}

\tableofcontents

\section{Introduction}

Electromagnetic form factors probe the distribution of the electrically charged
particles within hadrons. In experiment, they can be measured at
small space-like momentum transfers, $-q^2\equiv Q^2$, via the $Q^2$ dependence
of elastic scattering of electrons off hadrons, for instance. Here
we are interested in the electromagnetic form factor of the pion,
$f_{\pi\pi}(Q^2)$, which has been measured with high accuracy in the regime of
small space-like momentum transfers by the NA7
collaboration~\cite{Amendolia:1986wj}. Measurements at larger values of $Q^2$
have been done at CEA/Cornell~\cite{Bebek:1977pe},
DESY~\cite{Ackermann:1977rp,Brauel:1979zk} (a reanalysis of the data
can be found in refs.~\cite{Tadevosyan:2007yd,Huber:2008id}) and more recently
at JLab~\cite{Volmer:2000ek,Horn:2006tm,Horn:2007ug}.

From the theoretical point of view, the form factor is accessible analytically
only in the region of large $Q^2$ via perturbation
theory\cite{Brodsky:1973kr,Brodsky:1974vy,Farrar:1979aw,Efremov:1978rn,
Efremov:1979qk}, in contrast to the small and intermediate $Q^2$ regime, the
realm of strong coupling. At the current level of precision the form factor
from experiment at intermediate momentum transfers is well described by a simple
monopole ansatz of the form
\begin{equation}
 \label{eq:VPD}
f_{\pi\pi}(Q^2) = \left( 1 + Q^2/M^2_{\rm pole} \right)^{-1}
\end{equation}
where $M_{\rm pole}$ is the pole mass. This model is motivated by the
vector pole dominance
hypothesis~\cite{Holladay:1956zz,Frazer:1959gy,Frazer:1960zzb} (VPD), assuming
that the peak of the $\rho$-meson at time-like momentum transfers dominates the
behaviour of the form factor. To bridge the gap between experimental data and
the large $Q^2$ regime, one usually has to rely on models or effective field
theory predictions.

To address the strongly coupled small $Q^2$ region a non-perturbative treatment
is needed. A possible effective field theory framework is provided by chiral
perturbation theory~\cite{Gasser:1983yg,Gasser:1984gg,Gasser:1984ux}
(ChPT) where the $Q^2$ dependence of form factors can be computed order for
order in the quark mass and the quark momenta. The form factor has been computed
to next-to leading order (NLO) in refs.~\cite{Gasser:1984gg,Gasser:1984ux} and
in refs.~\cite{Bijnens:1998fm,Bijnens:2002hp}  to next-to-next-to leading order
(NNLO) in $SU(2)$ and $SU(3)$ ChPT, respectively. However, the expansion
includes unknown low energy constants (LECs) that need to be fixed to enable
predictions. In addition, the energy range to which ChPT at a given order is
applicable is not known {\it a priory} and thus has to be tested by comparison
to data. Concern regarding the effective field theory framework when applied to
electromagnetic form factors can be raised in regard of the tree-level coupling
of the photon to vector degrees of freedom. Extensions to ChPT including
vector degrees of freedom explicitly have been
formulated~\cite{Ecker:1988te,Ecker:1989yg} and the expression for the
electromagnetic form factor of the pion has been computed~\cite{Rosell:2004mn}.

Numerical simulations of lattice QCD offer a method to calculate QCD
observables from first principles and have been successfully applied to a number
of phenomenologically relevant quantities (for a compilation see
refs.~\cite{Colangelo:2010et,Laiho:2009eu}). Unfortunately, quantities related
to the structure of hadrons are more difficult. Particularly problematic are form
factors of baryons where the control of systematic effects such as excited state
contributions~\cite{Capitani:2012gj,Green:2012ud,Capitani:2012ef,
Bhattacharya:2013ava} and/or finite size effects~\cite{Yamazaki:2009zq} is
challenging (for recent reviews see
refs.~\cite{Lin:2012ev,Walker-Loud:2013yua}). This is not surprising
regarding the fact that even for $f_{\pi\pi}(Q^2)$, which is conceptually much
simpler, reliable continuum extrapolated results at the physical point have become
available only recently~\footnote{Note that the form factor
can also be calculated in the framework of Dyson-Schwinger equations
(see ref.~\cite{Chang:2013nia} and references therein).}.

Lattice measurements of $f_{\pi\pi}(Q^2)$ have been started in the late
80's~\cite{Martinelli:1987bh,Draper:1988bp} and initially been done neglecting
the effect of virtual quark
loops~\cite{vanderHeide:2003ip,vanderHeide:2003kh,AbdelRehim:2004sp,
Bonnet:2004fr,Capitani:2005ce,Hedditch:2007ex}. The aim of this review is to
provide an overview over the available lattice results for $f_{\pi\pi}(Q^2)$ and
the associated charge radius, with the focus on measurements including
dynamical quarks, started with the initial measurements reported in
ref.~\cite{Bonnet:2004fr}. For an easy assessment of the control over systematic
effects we will use a three-staged sign code in the spirit of
ref.~\cite{Colangelo:2010et}. We will start with a brief discussion of the
extraction of $f_{\pi\pi}$ and the associated theoretical developments and
challenges.

\section{Lattice computation of the form factor}
\label{sec:computat}

\subsection{Extraction of the form factor}
\label{sec:sys-eff}

The electromagnetic form factor of the pion is defined as
\begin{equation}
 \label{eq:fpipi_def}
 \left\langle\pi^+(\vec{p}_f)|V_\mu
   |\pi^+(\vec{p}_i)\right\rangle = (p_f+p_i)_\mu\,f_{\pi\pi}(Q^2)\,,
\end{equation}
where $V_\mu$ is the vector current, given by
\begin{equation}
 \label{eq:vcurr}
 V_\mu = \left\{ \begin{array}{ll} \displaystyle
 \frac{2}{3}\bar{u}\gamma_\mu u
  -\frac{1}{3}\bar{d}\gamma_\mu d & \textnormal{for} \quad N_f=2 \;;
\vspace*{2mm} \\
 \displaystyle \frac{2}{3}\bar{u}\gamma_\mu u
  -\frac{1}{3}\bar{d}\gamma_\mu d
  -\frac{1}{3}\bar{s}\gamma_\mu s \quad & \textnormal{for} \quad N_f=2+1
\end{array} \right.
\end{equation}
and $Q^2\equiv -(p_f-p_i)^2$ is the four-momentum transfer. At small momentum
transfers the $Q^2$ dependence of the form factor defines the charge radius
$\rpi$ and the curvature $c_V$ via
\begin{equation}
 \label{eq:ffexpans}
 f_{\pi\pi}(Q^2) = 1 - \frac{\rpi}{6} \; Q^2 - c_V \: Q^4 + \ldots \;. 
\end{equation}
Starting from the VPD model, eq.~\refc{eq:VPD}, the charge radius and the
curvature are given by $\rpi=6/M^2_{\rm pole}$ and $c_V = 1/M^4_{\rm pole}$.

The matrix element on the left hand side of eq.~\ref{eq:fpipi_def} at
space-like momentum transfers (time-like momentum transfers will
be mentioned in section~\ref{sec:summary}) can be extracted from the asymptotic
time-dependence of a suitable combination of Euclidean two and three-point
functions~\cite{Martinelli:1987bh,vanderHeide:2003ip,vanderHeide:2003kh,
AbdelRehim:2004sp,Bonnet:2004fr} in the limit of infinite temporal
separation between pion creation and annihilation operators and the current
insertion. At finite temporal extents one is left with contaminations from
excited states. This is particularly problematic at large momentum transfers
where the signal-to-noise ratio deteriorates exponentially with Euclidean
time~\cite{DellaMorte:2012xc}. One possibility to reduce the problem is to use
suitable ratios where some of the excited state contributions
cancel~\cite{Draper:1988bp,Capitani:2005ce,Hedditch:2007ex,
Bonnet:2004fr,Brommel:2006ww,Boyle:2007wg,Hsu:2007ai,
Boyle:2008yd,Frezzotti:2008dr,Aoki:2009qn,Nguyen:2011ek}. In practice, there are
two different domains where the extraction of the matrix element demands
different computational methods:

\begin{enumerate}

 \item {\bf Large momentum transfers:} \\
 For lattices with periodic boundary conditions, momenta of hadrons have to be
introduced by Fourier transformation, which in a finite volume leads to discrete
momenta. What we denote as the regime of large momentum transfers starts
around the smallest non-zero momentum transfer that can be achieved in this way
(e.g. for a typical lattice with $m_\pi L=4$ and $m_\pi=300$~MeV $Q^2_{\rm
min}=0.155$~GeV$^2$). In this regime the propagating pions obtain large momenta
so that the region where excited states are negligible is dominated by noise.

The standard method to overcome such a problem is to use suitable source and
sink operators that increases the overlap with the groundstate, thereby leading
to reduced excited state contaminations, usually at the cost of a slight growths
in the noise. This can be achieved by smearing source and sink
operators~\cite{Gusken:1989ad,Alexandrou:1990dq,Allton:1993wc}, with or without
blocked links~\cite{Teper:1987wt,Hasenfratz:2001hp} in the smearing kernel,
leading to a spatially extended source which mimicks the finite extent of the
quark bound state. A conventional implementation of this method allows to
extract the matrix element up to
$Q^2\approx3-4$~GeV$^2$~\cite{Bonnet:2004fr,Brommel:2006ww}. A combination with
all-to-all propagator methods offers the potential forfurther error
reduction~\cite{Aoki:2009qn}. To be able to go to even larger momentum transfers
it is necessary to use a smearing designed to improve the overlap with a boosted
state~\cite{Lin:2011sa,DellaMorte:2012xc}. In the first test,  where the
smearing parameters have been retuned for each momentum, momentum transfers as
large as 7~GeV$^2$ could be reached~\cite{Lin:2011sa}. Changing the kernel to an
anisotropic one has proven to reduce the error bars for pion two-point
functions~\cite{DellaMorte:2012xc}, while maintaining strong overlap with the
groundstate and offers the possibility for further improvement. Other methods
to reduce excited state contaminations are provided by
variational~\cite{Michael:1982gb,Luscher:1990ck,Blossier:2009kd} and/or
summed operator
insertion~\cite{Maiani:1987by,Gusken:1989ad,Bulava:2011yz,Capitani:2012gj}
techniques. However, for very large momentum transfers, larger than the inverse
of the lattice spacing squared (typically $a^{-2}\sim 6$ to 16~GeV$^2$), large
discretisation effects can appear which demand a careful extrapolation to the
continuum. To overcome this problem a computationally demanding step-scaling
procedure has been proposed in ref.~\cite{Lin:2011sa}, but has not been put to
a test so far.

 \item {\bf Small momentum transfers:} \\
 The regime below the minimal momentum transfer available by Fourier
transformation can be accessed by using {\it partially twisted} boundary
conditions~\cite{Bedaque:2004kc,Sachrajda:2004mi,deDivitiis:2004kq,Flynn:2005in}
which allow to extract the form factor at arbitrary momentum
transfers~\cite{Boyle:2007wg}. The phrase {\it partially} refers to the fact
that only the boundary conditions for the quark fields in the computation of the
propagators are changed. This procedure introduces an additional finite size
effect which can be shown to decrease exponentially with the volume for pionic
matrix elements~\cite{Sachrajda:2004mi}. Since in this regime the exponential
decay of the signal-to-noise ratio is usually small enough to provide a signal
for the whole range between source and sink, the use of stochastic
estimators~\cite{Bernardson:1993he,Dong:1993pk,Foster:1998vw,
McNeile:2006bz}, leading to a substantial error reduction for correlation
functions with propagating pions~\cite{Boyle:2008rh}, enables the extraction
of the form factor with high accuracy.

\end{enumerate}

The main systematic effects that need to be controlled in the process of
extracting the matrix element on a given ensemble are: (i) contamination from
excited states; (ii) renormalisation of the vector current; (iii) full
$\mathcal{O}(a)$-improvement. Whether the last two of these systematic effects
apply depends on the details of the computation. Other systematic
effects connected with the extrapolation to the physical point will be
discussed in the next section. Concerning the contaminations from excited
states there are two types~\cite{Brommel:2006ww,Brandt:2013dua}: (a) a
contribution independent of the time of the current insertion, decreasing
exponentially with the temporal separation between creation and annihilation
operator; (b) a contribution decreasing exponentially with the temporal
difference between those and the current insertion. The latter can easily be
avoided by extracting the matrix element from the region where only a single
state contributes to the decay of 2 and 3-point functions. Contaminations of
type (a) are more difficult to remove, but are strongly suppressed and
can typically be neglected. However, with increasing precision the
importance of the effect is enhanced and can possibly be of the order of the
error bars in todays simulations~\cite{Brandt:2013dua}.

\subsection{Extrapolation to the physical point}
\label{sec:phextrapol}

Even though most of the collaborations today have lattices very close to, or
even at the physical point at their disposal, current measurements of the form
factor are still restricted to unphysically large pion masses. The calculation
of the form factor at the physical point thus demands a chiral extrapolation on
top of the usual extrapolations to the continuum (lattice spacing $a$ $\to0$)
and infinite volume. 

The main uncertainty in the chiral extrapolation originates from the
unknown exact functional form valid for a given range of pion masses. Here ChPT
can provide guidance~\cite{Bijnens:1998fm,Bijnens:2002hp} even though the range
of validity at a given order is also not known {\it a priory}. The full $SU(2)$
expression at NNLO is given in terms of the pion mass $m_\pi$ and the pion decay
constant $F_{\pi}$ and contains five free parameters. Four of them remain in the
expression for the charge radius. In contrast, at NLO it contains only one free
parameter. At NLO the expression is basically linear in $Q^2$
(up to a mild $Q^2$-dependence in the function $J(Q^2)$~\cite{Bijnens:1998fm}),
so that its validity will be limited to the region where the curvature of the
form factor is negligible, while at NNLO it contains terms up to
$\mathcal{O}(Q^6)$. In practice, it turns out that for the pion masses in reach
the data cannot be described consistently with ChPT to
NLO~\cite{Frezzotti:2008dr,Aoki:2009qn,Nguyen:2011ek,Brandt:2011jk,
Brandt:2013dua}. On the other hand, the large number of free parameters at NNLO
are typically not sufficiently constrained by the curvature and the pion mass
dependence of the form factor
alone~\cite{Frezzotti:2008dr,Aoki:2009qn,Nguyen:2011ek,Brandt:2013dua}. To
overcome this problem, most collaborations have used the appearance of several
LECs in the expressions for different quantities for a joined chiral
extrapolation, typically including the pion decay constant $F_\pi$ and the pion
mass $m_\pi$~\cite{Aoki:2009qn,Frezzotti:2008dr,Nguyen:2011ek,Brandt:2013dua}.
The region of validity of the effective theory can potentially be increased when
vector degrees of freedom are added, however, the associated formula for the
form factor~\cite{Rosell:2004mn} also contains a number of free parameters and
has so far not been compared to lattice data.

Apart from ChPT one can also use other ans\"atze for the chiral extrapolation.
One example is to use a chiral extrapolation for the pole
mass~\cite{Brommel:2006ww}, defined via eq.~\refc{eq:VPD}. One should, however,
keep in mind that such a chiral extrapolation depends on the validity of the VPD
hypothesis. Another possibility is to use a polynomial extrapolation for the
charge radius~\cite{Frezzotti:2008dr,Brandt:2013dua} which usually leads to
reasonable agreement with the extrapolation via ChPT.

For the extrapolation to infinite volume it is often too expensive to perform
simulations with more than one volume for each simulation point, so that one
has to rely on alternative methods. One possibility is to do a finite volume
study only for a few ensembles and to generalise the findings to the full set of
simulation points. The extrapolation to the infinite volume
limit can also be done in combination with the chiral extrapolation in the
framework of ChPT. The expressions are given for Fourier momenta in
ref.~\cite{Borasoy:2004zf} and twisted boundary conditions in
refs.~\cite{Jiang:2006gna,Jiang:2008te}~\footnote{For $F_\pi$ and $m_\pi$
similar calculations have been done in ref.~\cite{Colangelo:2005gd}.}. Finite
volume effects for the form factor have also been studied for the
$\epsilon$-regime~\cite{Gasser:1986vb} at leading order~\cite{Fukaya:2012dla}.
This allows to extract the form factor on very small lattices, but relies on
the assumption that higher order corrections are small.

The extrapolation to the continuum is conventionally done using a polynomial in
powers of the lattice spacing $a$. Assuming that the theory is fully
$\mathcal{O}(a)$ improved, this means that the $Q^2$-expansion of the form
factor contains terms of the form $a^{2n}Q^{2m}$, where $n,m\in \mathbb{N}$.
The terms $a^2 Q^2$ and $a^2 Q^4$ represent lattice artefacts for $\rpi$ and
$c_V$, respectively.

\section{Compilation of results for form factor and charge radius}

\begin{figure}
\begin{center}
\includegraphics[angle=-90, width=.65\textwidth]{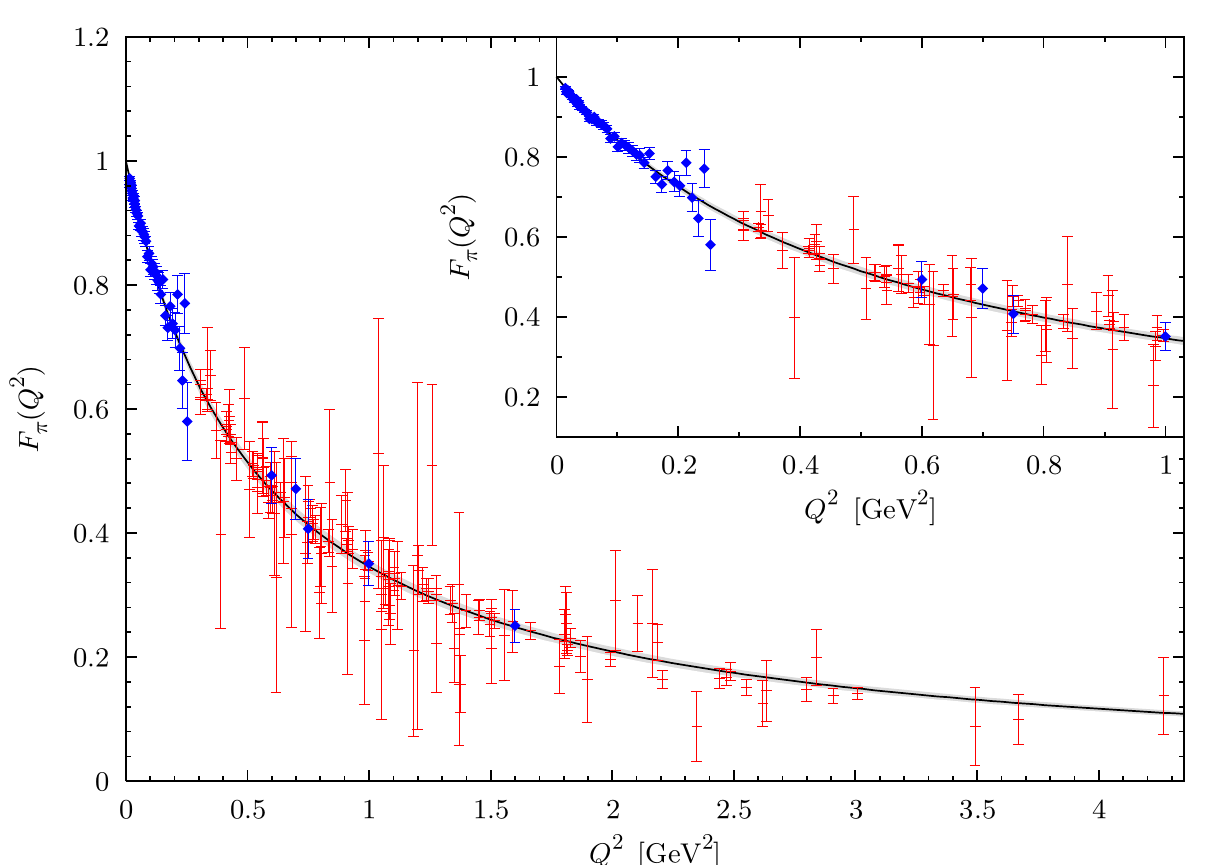}
\vspace*{-2mm}
\end{center}
\caption{Results for the electromagnetic form factor of the pion in the
regime of large momentum transfers from the QCDSF/UKQCD
collaboration~\cite{Brommel:2006ww} in comparison to results from
experiment~\cite{Amendolia:1986wj,Brauel:1979zk,Volmer:2000ek}.} 
\label{fig:QCDSF_FF}
\end{figure}

\subsection{Results for the form factor}

A number of collaborations have computed the form factor in the large momentum
region with momentum transfers up to 4~GeV$^2$ using
Fourier momenta~\cite{Bonnet:2004fr,Brommel:2006ww,Hsu:2007ai,Boyle:2007wg,
Boyle:2008yd,Aoki:2009qn,Nguyen:2011ek,Lin:2011sa,Fukaya:2012dla} and twisted
boundary conditions~\cite{Frezzotti:2008dr}. All results show good agreement
with a one-parameter pole ansatz, eq.~\refc{eq:VPD}. Newer studies with
increasing accuracy, however, observe deviations from the single
pole form~\cite{Aoki:2009qn,Fukaya:2012dla}. So far, most of the studies in the
large $Q^2$-regime have only been done on a single lattice spacing, so that
usually a continuum extrapolation is missing. The only exception is the study
by the QCDSF/UKQCD collaboration~\cite{Brommel:2006ww}, with results up to
$Q^2\lesssim4.5$~GeV$^2$. They have obtained results at five different
lattice spacings and various pion masses in the range of 400 to 1000~MeV,
enabling a chiral and continuum extrapolation. This extrapolation, however, has
been done for the pole mass directly and is thus valid only within the VPD
model. As shown in figure~\ref{fig:QCDSF_FF} the results are in good
agreement with the experimental data within the error bars.

\begin{figure}
\begin{center}
\vspace*{-3mm}
\includegraphics[width=.8\textwidth]{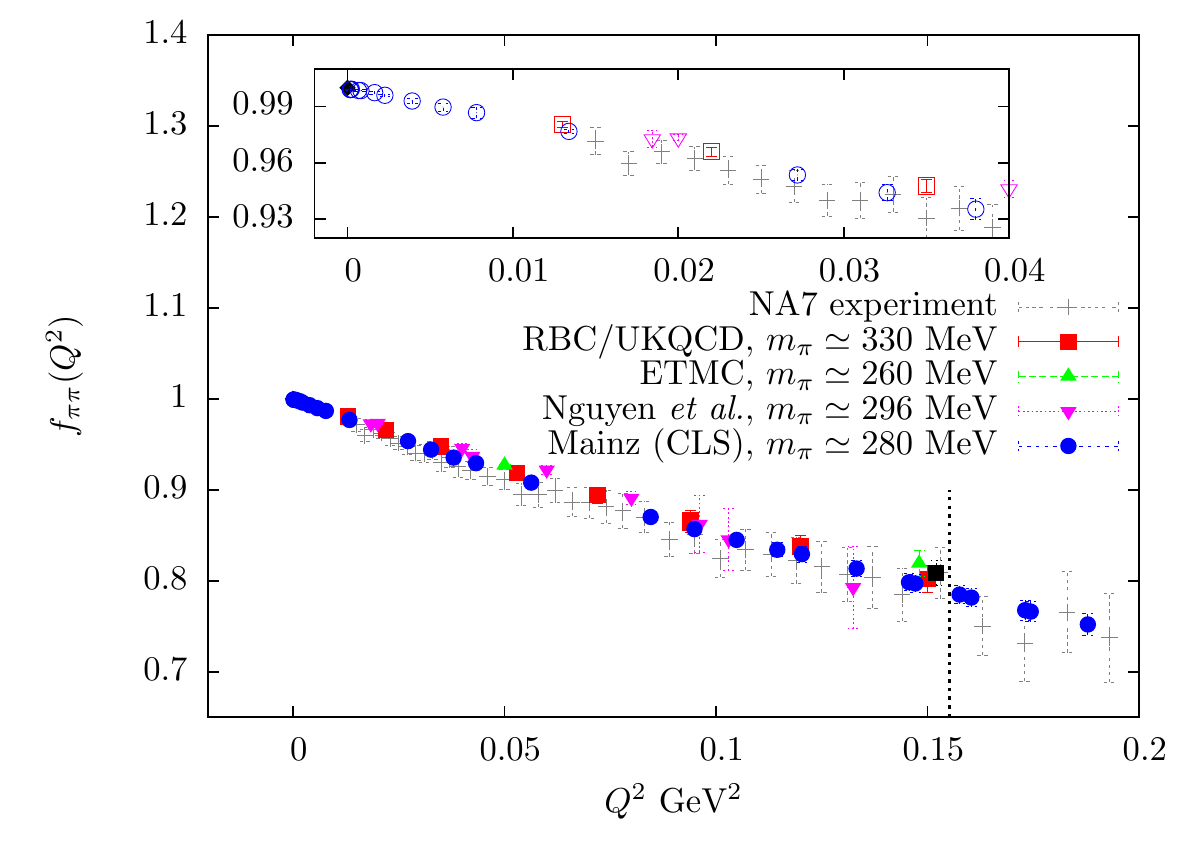}
\vspace*{-6mm}
\end{center}
\caption{Compilation of lattice results for the pion form
factor~\cite{Boyle:2008yd,Frezzotti:2008dr,Nguyen:2011ek,Brandt:2013dua} in the
region of small $Q^2$ in comparison to the experimental results of the NA7
collaboration~\cite{Amendolia:1986wj}. The black data point is the result of the
RBC/UKQCD collaboration at the smallest available $Q^2$ from Fourier momenta and
the black dashed line indicates $Q^2_{\rm min}=0.155$~GeV$^2$ the associated
value for an average lattice with $m_\pi=300$~MeV and $m_\pi\;L=4$. The inset
highlights the very small $Q^2$-region, displaying the potential of partially
twisted boundary conditions in the approach of $Q^2\to0$.}
\label{fig:FF_compil_smallqsq}
\end{figure}

In the small $Q^2$ regime, the form factor can be extracted with high precision
and calculations have been done by the
RBC/UKQCD~\cite{Boyle:2007wg,Boyle:2008yd},
ETM~\cite{Frezzotti:2008dr} and PACS-CS~\cite{Nguyen:2011ek} collaborations
and recently by the Mainz group~\cite{Brandt:2013dua}. A compilation of results
for the form factor in this regime, including the smallest pion mass of each
collaboration, is shown in figure~\ref{fig:FF_compil_smallqsq}. Even though at
unphysically large pion masses, explaining the tendency towards larger
$f_{\pi\pi}$ values, the data shows remarkable agreement with the experimental
results.

\vspace*{-2mm}
\subsection{Extraction of the charge radius}
\label{sec:chrad-ext}

\begin{figure}[t]
\begin{center}
\vspace*{-3mm}
\includegraphics[width=.8\textwidth]{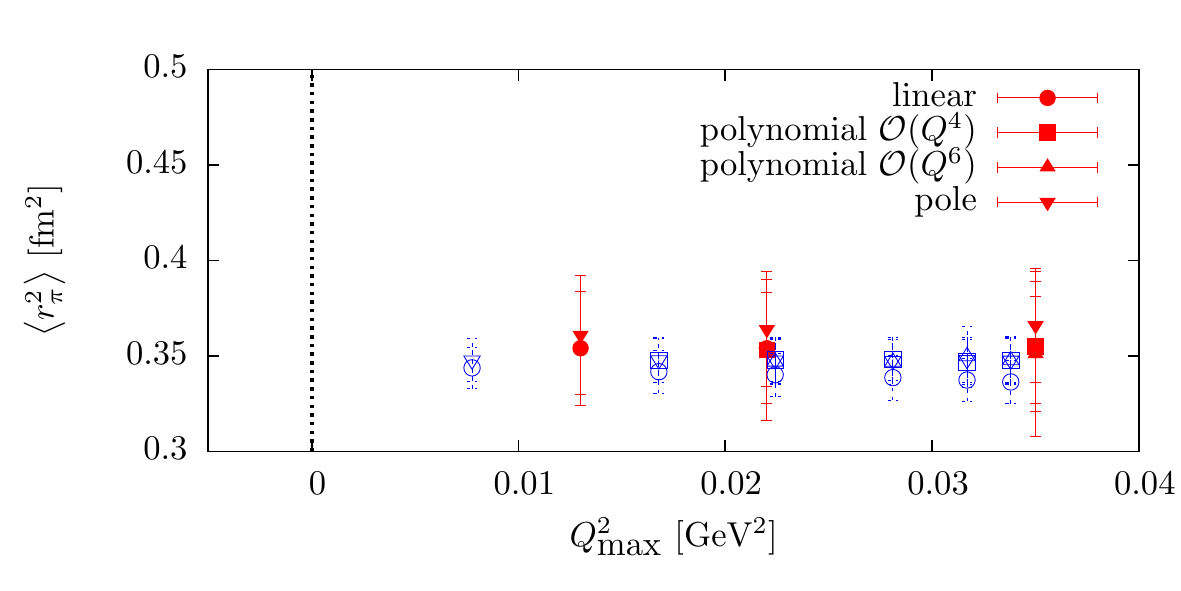}
\vspace*{-5mm}
\end{center}
\caption{Results for the form factor from the RBC/UKQCD
collaboration~\cite{Boyle:2008yd} (red filled symbols) at a pion mass of 330~MeV
and the Mainz group~\cite{Brandt:2013dua} (blue open symbols) at a pion mass of
325~MeV extracted from a fit to the form given in the legend.} 
\label{fig:RPI_extract}
\end{figure}

The electromagnetic charge radius of the pion, $\rpi$, is defined by the
expansion of the form factor given in eq.~\refc{eq:ffexpans}. Its extraction
suffers from an inherent model dependence, associated with the particular ansatz
for $f_{\pi\pi}(Q^2)$, unless data at very small momentum transfers is
available. A comparison of the results for $\rpi$ at $m_\pi\approx330$~MeV,
obtained from different functional forms including all data for the form factor
up to the $Q^2_\text{max}$ value indicated on the horizontal axis by
the RBC/UKQCD collaboration~\cite{Boyle:2008yd} and of the Mainz
group~\cite{Brandt:2013dua}, is shown in figure~\ref{fig:RPI_extract}. The
equivalence of the results from different fits below $Q^2\approx0.025$~GeV$^2$
indicates that, at the given accuracy, contributions from terms beyond the
linear term are negligible. At larger momentum transfers the result from the
linear fit of the Mainz group start to deviate from the other functional forms.
This is not visible for the results from RBC/UKQCD, albeit possibly lost in the
larger statistical uncertainties. Interestingly, the one-parameter pole fit
agrees well with the unconstrained polynomials up to $Q^2=0.04$~GeV$^2$.
However, this is not guaranteed to persist when data at larger $Q^2$ are
included. In fact, data from RBC/UKQCD and also from the Mainz group indicate
that the result from the one-parameter pole fit increase in this case. A similar
tendency has been observed by the JLQCD/TWQCD collaboration~\cite{Aoki:2009qn}
where the pole fits give smaller $\rpi$ values when polynomial terms are added.
In principle, a similar analysis should be done for the experimental data.
However, the relatively large error bars in the low $Q^2$-region and the
uncertainty of the normalisation at $Q^2=0$ of the data from the
NA7-collaboration~\cite{Amendolia:1986wj} makes it difficult to obtain
significant results for $Q^2_{\rm max}\leq0.1$ only.

\begin{figure}[t]
\begin{center}
\vspace*{-3mm}
\includegraphics[width=.8\textwidth]{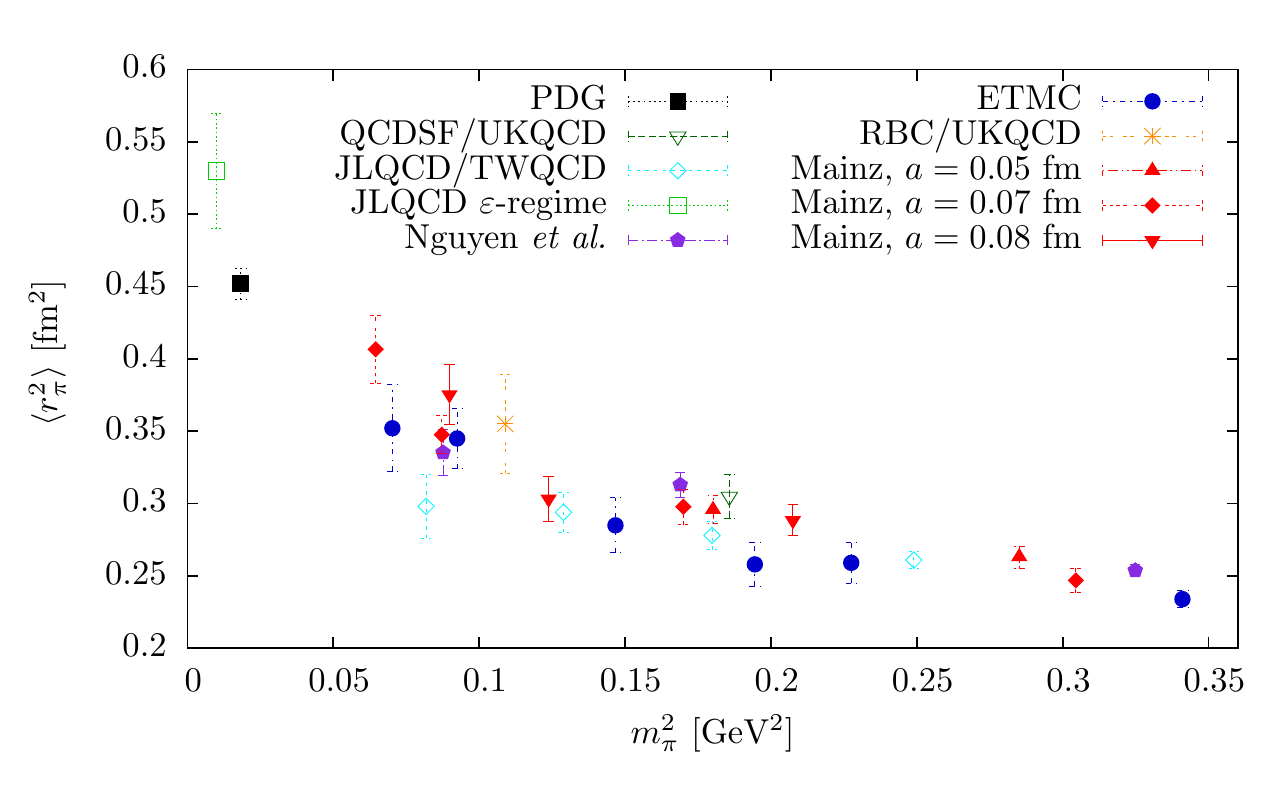}
\vspace*{-5mm}
\end{center}
\caption{Compilation of results for the charge radius from lattice
QCD~\cite{Brommel:2006ww,Aoki:2009qn,Fukaya:2012dla,Nguyen:2011ek,
Frezzotti:2008dr,Boyle:2008yd,Brandt:2013dua} versus the squared pion mass in
physical units (here $m_\pi\leq600$~MeV) and the
experimental average quoted by the PDG~\cite{Nakamura:2010zzi}. Open symbols
denote results extracted from form factor data with Fourier momenta only, while
filled symbols represent results including data from twisted boundary
conditions. The results for the QCDSF/UKQCD collaboration have been rescaled
using the updated data for the lattice spacing (Sommer scale) obtained in
ref.~\cite{Bali:2012qs}.}
\label{fig:chrad-results}
\end{figure}

Figure~\ref{fig:chrad-results} contains a collection of results for the charge
radius in the region of pion masses up to $600$~MeV. Open symbols indicate that
the result has been obtained from Fourier momenta only. The plot illustrates the
good overall agreement between different collaborations. In particular, the
results from the QCDSF/UKQCD collaboration rescaled with the updated lattice
spacing from ref.~\cite{Bali:2012qs} also agree well with the other lattice
determinations.

\section{Results at the physical point}
\label{sec:PRES}

\begin{table}
\centering
\small
\vspace*{-3mm}
\begin{tabular}{l|l|l|l}
 Systematic effect & abbrev. & criteria for \sysfullcontrol & criteria for
 \syshalfcontrol \\
\hline
\hline
 Chiral & ChE & $m_{\pi,{\rm min}} < 250$~MeV & $m_{\pi,{\rm min}} \leq
400$~MeV \\
 extrapolation & & & \\
\hline
 Continuum & CoE & 3 or more & 2 lattice spacings$^{(i)}$\\
 extrapolation & & lattice spacings$^{(i)}$ & \\
\hline
 Finite volume & FVE & $m_\pi\;L\gtrsim3.7$ & $m_\pi\;L\gtrsim3$ \\
 effects & & \&/or 2 volumes$^{(ii)}$ & \\
\hline
 Model dependence & MD & at least 2 points with & 
 a detailed comparison of \\
 in $\rpi$ extraction & & $Q^2<0.03$~GeV$^2$ & several fit functions
\\
\hline
\hline
\end{tabular}
\caption{List of systematic effects evaluated for the extraction of the charge
radius. If the criteria above have not been fulfilled we have attributed a
`\sysnocontrol' sign. $^{(i)}$ We also adopt the additional criteria from the
update~\cite{FLAG_new} of the FLAG review: $a^2_{\rm max}/a^2_{\rm min}\geq2$,
$D(a_{\rm min})\leq2$~\% and $\delta(a_{\rm min})\leq1$ for a
`\sysfullcontrol'; $a^2_{\rm max}/a^2_{\rm min}\geq1.4$,
$D(a_{\rm min})\leq10$~\% and $\delta(a_{\rm min})\leq2$ for a `\syshalfcontrol'
(see ref.~\cite{FLAG_new}, section 2, for notation and details). $^{(ii)}$ The
two volumes have to be at fixed other parameters and in the case of
$m_\pi\;L\gtrsim3.7$ they can be replaced by estimating the effect in ChPT
instead.}
\label{tab:syseffects}
\vspace*{2mm}
\begin{tabular}{lc|l|cccc|l}
 Collaboration & \hspace*{-4mm}$N_f$ & ChE mth. & ChE & CoE & FVE & MD & $\rpi$
[fm$^2$] \\
\hline
\hline
 & & & & & & & \vspace*{-3mm} \\
QCDSF/UKQCD~\cite{Brommel:2006ww} & \hspace*{-4mm}2 & $M_{\rm pole}$ &
\sysnocontrol &
\sysfullcontrol & \syshalfcontrol & \sysnocontrol &
0.509(22)(74) \\
ETMC~\cite{Frezzotti:2008dr} & \hspace*{-4mm}2 & NNLO $f_{\pi\pi}$ & \syshalfcontrol &
\syshalfcontrol &
\syshalfcontrol & \syshalfcontrol & 0.456(30)(24) \\
JLQCD/TWQCD~\cite{Aoki:2009qn} & \hspace*{-4mm}2 & NNLO $f_{\pi\pi}$ & \syshalfcontrol &
\sysnocontrol &
\sysnocontrol & \syshalfcontrol & 0.409(23)(37) \\
Mainz~\cite{Brandt:2013dua} (CLS) & \hspace*{-4mm}2 & NNLO $\rpi$ & \syshalfcontrol &
\sysfullcontrol &
\sysfullcontrol & \sysfullcontrol & 0.481(34)(13) \\
\hline
 & & & & & & & \vspace*{-3mm} \\
RBC/UKQCD~\cite{Boyle:2008yd} & \hspace*{-4mm}2+1 & NLO & \syshalfcontrol &
\sysnocontrol &
\sysfullcontrol & \sysfullcontrol & 0.418(31) \\
Ngyuen {\it et al$\:$}~\cite{Nguyen:2011ek} & \hspace*{-4mm}2+1 & NNLO $\rpi$ &
\syshalfcontrol &
\sysnocontrol & \sysfullcontrol & \sysfullcontrol & 0.441(46) \\
\hline
 & & & & & & & \vspace*{-3mm} \\
PDG~\cite{Nakamura:2010zzi} & & --- & & & & & 0.452(11) \\
Amendolia {\it et al$\:$}~\cite{Amendolia:1986wj} & & --- & & & & & 0.439(8) \\
BCT~\cite{Bijnens:1998fm} & & --- & & & & & 0.437(16) \\
\hline
\hline
\end{tabular}
\caption{Collection of results for the electromagnetic charge radius of the
pion from Lattice QCD with
two~\cite{Brommel:2006ww,Frezzotti:2008dr,Aoki:2009qn,Brandt:2013dua} and
three~\cite{Boyle:2008yd,Nguyen:2011ek} dynamical quark flavours,
experiment~\cite{Nakamura:2010zzi,Amendolia:1986wj} and an analysis of the
experimental data using ChPT at NNLO. `ChE mth.' denotes the method used for
the chiral extrapolation and $N_f$ stands for the number of dynamical quarks
used in the simulations ($2+1$ means two degenerate light and a heavier strange
quark). The result for the QCDSF/UKQCD collaboration is the one which is
obtained from the new scale determination~\cite{Bali:2012qs}. The original
result~\cite{Brommel:2006ww} was $\rpi=0.442(19)(64)$~fm$^2$. For the
associated chiral extrapolation we have given the \sysnocontrol sign, following
the criteria listed above, in contrast to the assessment of the FLAG review
(`green' flag~\cite{Colangelo:2010et,FLAG_new}), since the minimal pion mass
included is about 428~MeV with the new lattice spacing.}
\label{tab:chrad}
\end{table}

After the discussion of the results at non-physical pion mass we will now
discuss the results for the form factor and charge radius that have been
extrapolated to the physical point. To provide an indication for the control
over the main systematic effects for non-experts regarding lattice QCD, we will
follow the strategy of the FLAG group~\cite{Colangelo:2010et} and introduce
a sign code, indicating how well a particular systematic effect is under control
in an individual measurement. The criteria and systematic effects considered for
the evaluation are collected in table~\ref{tab:syseffects} and we use the signs
\sysfullcontrol, \syshalfcontrol and \sysnocontrol for full, partial and no
control over the effect, respectively. The first three systematic effects and
criteria have been adopted from the next update~\cite{FLAG_new} of the FLAG
review~\cite{Colangelo:2010et} which already contains an evaluation of the
results for the charge radius (note that in the LEC section the criteria for
the chiral extrapolation are those given in
table~\ref{tab:syseffects}~\footnote{I would like to thank Stephan D\"urr for
pointing this out.} and are different to the rest of ref.~\cite{FLAG_new}). For
those three criteria we agree with the assessments in the FLAG review (except
for the evaluation concerning the chiral extrapolation of the QCDSF result). At
this point it is important to stress that the ChPT formulae at NNLO for $\rpi$
(or $f_{\pi\pi}$), $F_\pi$ and $m_\pi$ still has 11 to 14 free parameters
(depending on the extrapolation strategy) so that some of the parameters still
need to be constrained to stabilise the fits. For this different strategies have
been pursued~\cite{Aoki:2009qn,Frezzotti:2008dr,Nguyen:2011ek,Brandt:2013dua},
thereby introducing another systematic effect concerning the chiral
extrapolation, whose impact is difficult to quantify.

We have added the model dependence of the extraction of the charge radius as
an additional systematic effect which is an issue for the extraction of charge
radii in general as long as there is a gap in the form factor data to $Q^2=0$.
In the case of $\rpi$ we have seen in the last section that the form factor is
basically linear below $Q^2\approx0.03$~GeV$^2$ with the present accuracy of
the data, so that the residual model dependence in the extraction of $\rpi$ is
negligible. We have also seen that a careful extraction
of $\rpi$ comparing different functional forms leads to results that are
compatible with the ones obtained from $Q^2$ values below 0.03~GeV$^2$ so that
in this case there is at least some control over the associated systematic
effect.

\begin{figure}[t]
\begin{center}
\vspace*{-3mm}
\includegraphics[width=.57\textwidth]{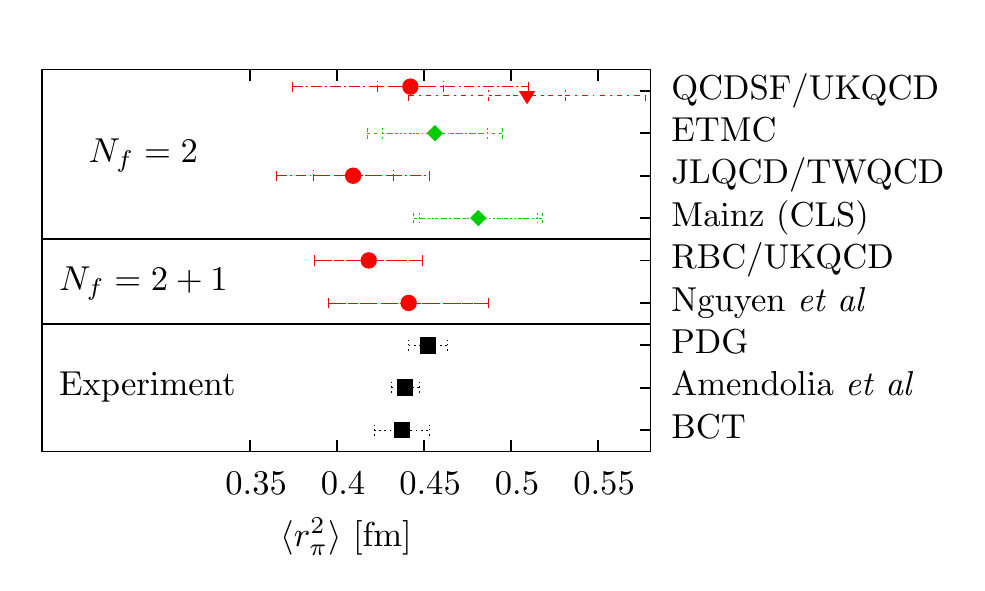}
\end{center}
\vspace*{-5mm}
\caption{Results for the pion charge radius at the physical point as given in
table~\ref{tab:chrad}. The two different results for the QCDSF/UKQCD
collaboration are the results obtained with $r_0=0.467$~fm (circle) and
$r_0=0.501$~fm (triangle) (see the caption of table~\ref{tab:chrad}).}
\label{fig:RPI}
\end{figure}

The available results for $\rpi$ at the physical point are listed in
table~\ref{tab:chrad}. In the assessment of systematic effects we have
assumed that all collaborations fully control the effects concerning the
extraction of the matrix element on a given ensemble (cf.
section~\ref{sec:sys-eff}). The table also includes
experimental results, the NA7 result~\cite{Amendolia:1986wj} and the PDG
average~\cite{Nakamura:2010zzi}, as well as a result from ChPT applied to the
experimental data~\cite{Bijnens:1998fm}. These results in principle share the
model dependence in the extraction of $\rpi$ since experimental data is only
available for $Q^2>0.015$~GeV$^2$~\footnote{The data from experiment at space
and time-like $q^2$ can also be used to derive bounds for the pion charge radius
that are potentially model independent (see
ref.~\cite{Ananthanarayan:2013dpa}).}. A scatter plot of the collection of
results 
is shown in figure~\ref{fig:RPI}. The measurements with at least partial control
over all systematic effects, the ones from the ETMC~\cite{Frezzotti:2008dr} and
the Mainz group~\cite{Brandt:2013dua}, are marked in green (diamonds). The plot
displays the overall agreement, even though the data shows a certain spread with
the more reliable results at the upper end. Note, that the remaining systematic
uncertainty is mostly due to the chiral extrapolation. Future lattice
calculations will have to improve on this to make a precise prediction for the
charge radius at the physical point.

\begin{figure}[t]
\begin{center}
\vspace*{-3mm}
\includegraphics[width=.8\textwidth]{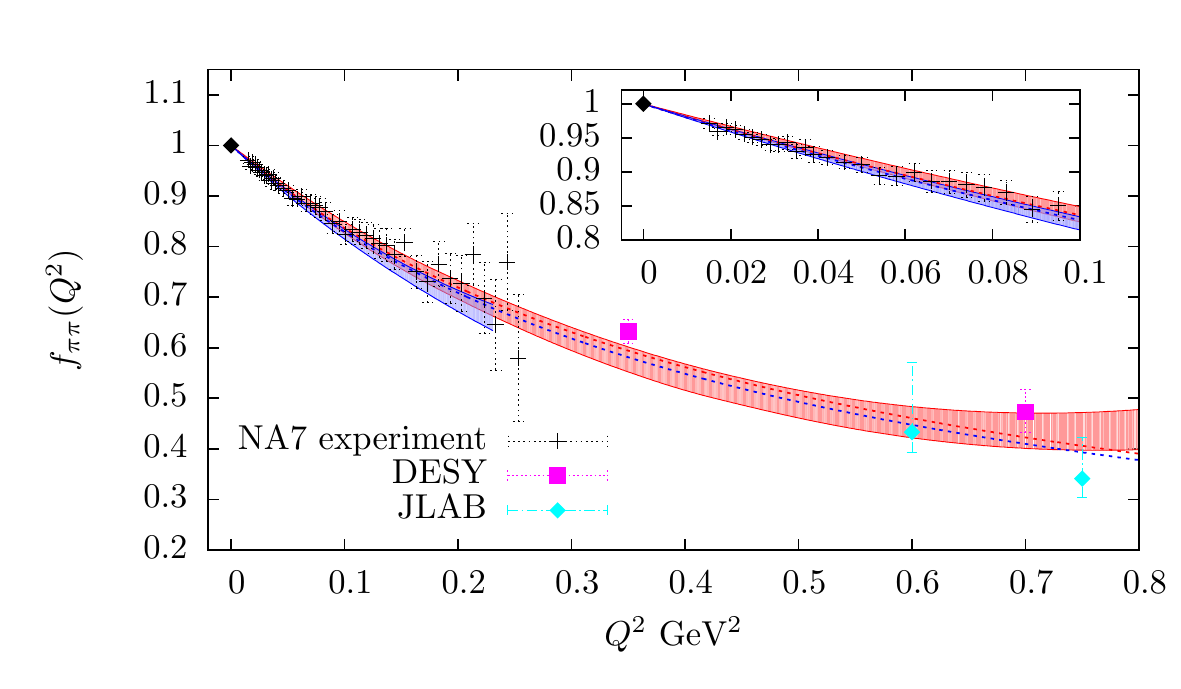}
\vspace*{-5mm}
\end{center}
\caption{Results for the $Q^2$-dependence of the form factor at the physical
point obtained from lattice QCD (red - higher curve: ETMC~\cite{Frezzotti:2008dr};
blue - lower curve: Mainz~\cite{Brandt:2013dua}) using ChPT at NNLO (coloured areas)
and the VPD model (dashed lines).}
\label{fig:FPIPI}
\end{figure}

The chiral extrapolation for the form factor is typically done in the framework
of ChPT, so that the $Q^2$-dependence of the form factor at the physical point
can naturally be represented by its ChPT expression using the LECs from the
chiral extrapolation. Figure~\ref{fig:FPIPI} shows $f_{\pi\pi}(Q^2)$ up to
$Q^2=0.8$~GeV$^2$ in ChPT at NNLO obtained by ETMC~\cite{Frezzotti:2008dr} and
the Mainz group~\cite{Brandt:2013dua}. Note, that the results of the Mainz group
only use results with $Q^2\leq0.077$~GeV$^2$ and the results from ETMC are not
extrapolated to the continuum. Both curves show agreement with the
experimental data. The plot also includes the $Q^2$-dependence (dashed lines)
obtained from the VPD model with $\rpi=6/M^2_{\rm pole}$ and the value for
$\rpi$ at the physical point.

For the curvature $c_V$ less results are available in the literature, the
only two being the ones from ETMC~\cite{Frezzotti:2008dr},
$c_V=3.37(31)(27)$~GeV$^{-4}$, and JLQCD/TWQCD~\cite{Aoki:2009qn},
$c_V=3.22(17)(36)$~GeV$^{-4}$. As indicated in table~\ref{tab:chrad} the only
one for which the main systematic effects are partially under control is the
ETMC result.

\section{Summary and perspectives}
\label{sec:summary}

This review contains a summary of the available measurements of the
electromagnetic form factor of the pion from numerical simulations in full
lattice QCD. A list of results of the charge radius has been given in
table~\ref{tab:chrad} and particular attention has been attributed to the
indication of the control over systematic effects. Owing to partially twisted
boundary conditions, lattice QCD has the unique opportunity to calculate the
form factor at arbitrarily small momentum transfers, which
allows for extracting the charge radius of the pion without residual model
dependence. The remaining dominant systematic uncertainty is due to the chiral
extrapolation, but results at the physical point, that will hopefully become
available in the near future (first results have been presented at this years
lattice conference by the MILC
collaboration~\cite{MILClattice}), will remove this uncertainty. An accurate
extraction of the charge radius, however, also demands full control over the
other systematic effects. In particular, lattice simulations are usually done
with degenerate light quarks and thus neglect the effects due to iso-spin
breaking (for a recent review see ref.~\cite{Portelli:2013jla}). The reliable
inclusion of these effects is one of the main tasks for the future.

The accurate extraction of the form factor at large momentum transfers is
computationally even more challenging due to the earlier loss of the signal in
the extraction of the matrix element. Clearly more work is needed to increase
the control over the systematic effects and to decrease the error bars in this
regime. Possible strategies have been discussed in section~\ref{sec:sys-eff}.

Time-like momentum transfers are not directly accessible in lattice simulations
(at least for non-transition form factors) but can be accessed indirectly for
the $Q^2$ range between the two-particle threshold ($2m_\pi$) and the inelastic
threshold ($4m_\pi$), following a recent proposal~\cite{Meyer:2011um}.
Even though the method is computationally demanding first results have been
presented recently~\cite{timelikeFF}.

{\bf Acknowledgements:} 
I am grateful to Andreas J\"uttner for carefully going through the manuscript
and illuminating discussions, Hartmut Wittig for numerous discussions on the
matter, Silvano Simula for providing access to the data for the form factor of
the ETM collaboration at the physical point and Jacques Bloch for information
about the Dyson-Schwinger measurements. 

\vspace*{-3mm}
\bibliographystyle{ws-ijmpe}
\bibliography{library}

\end{document}